# scientific reports

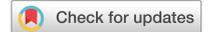

OPEN

# Investigation of physical and electrical properties of a suboxide layer at Si/Si-hexafluoride interface

**Seref Kalem**

The silicon suboxide SiOx (x < 2.0) offers promising industrial application possibilities ranging from electrodes in lithium-ion batteries, which are used widely in electrical vehicles and portable devices to sensing applications. Therefore, a low cost, environmental friendly and high performance silicon oxide materials are required for an appropriate operation of any electronic gadget. In this work, we report on the physical and electrical properties of a suboxide layer of up to 1 µm, which was grown on silicon during the formation of a dielectric layer, namely the ammonium silicon hexafluoride. It is a stable oxide exhibiting light emission from 400 to 1700 nm offering scalable and cost-effective large area processing capability. The measurement results reveal interesting properties, which are required to be understood clearly before proceeding with any suitable application. The results have been analyzed using state-of-the-art physical and electrical characterization techniques such as ellipsometry, AFM, SEM, FTIR, photoluminescence lifetime and resistive switching measurements to determine structural, optical and electrical properties. At 300 K the carrier lifetime measurements reveal the lifetime values ranging from about few tens of picosecond up to 4500 picoseconds. Scanning probe analysis indicate a surface roughness of about 30 Å. Resistive memory forming was observed also in these layers at relatively low power thresholds. We provide a comprehensive description of the physical and electrical properties in order to clarify the origin of the observed features. The wavelength dependent real $\varepsilon_1(\omega)$ and the imaginary $\varepsilon_2(\omega)$ dielectric functions provided useful insights on optical properties. A lookout is given for the possible applications of this special SiOx dielectric oxide layer.

In recent years, silicon suboxide has attracted a great deal of interest due to rich physical and electrical properties with suitability for commercial development[1–4]. Among others, its potential for possible application as anode material in lithium-ion batteries has received interest in energy industry for the development of high performance rechargeable batteries[5,6]. Silicon suboxide appears to be a strong candidate in replacing metallic lithium anodes, which hindered their application due to parasitic reactions and short circuiting hazards[7]. Therefore, the exploration of the potential of silicon based anodes might offer solution to the problem of electrode degradation. Actually, a number of leading companies are considering the integration of silicon anodes to their production lines[8]. According to the forecasts made on the silicon anode market, multi-billion USD worth of business appears to be plausible[9,10]. A detailed description of the road map on silicon anodes in lithium-ion batteries has been provided very recently[11].

The technical aspect of our research work is based on the investigation of physical and electrical properties of silicon suboxide oxide layer grown during the ammonium silicon hexafluoride dielectric formation process[12]. The silicon suboxide in this process played the role of a oxide layer for the growth of the dielectric. The removal of deionized water soluble dielectrics leaves the silicon suboxide intact on the surface of the silicon wafer. The oxide so grown has been subjected to a number of state-of-the-art characterization techniques for measurement of electrical, structural and physical properties. The scientific aspect of this work is related to understanding of the interface between the Si-hexafluoride and silicon wafer. The technological significance is based on the scalable production of sub-oxides using cost-effective wet-chemical processing method.

Department of Electrical and Electronics Engineering, Faculty of Engineering and Natural Sciences, Bahcesehir University, 34353 Besiktas, Istanbul, Turkey. email: seref.kalem@bau.edu.tr

Localized vibrational modes analysis reavealed the presence of Si–TO and Si–O–Si stretching bands as well as TO–LO mode couplings. The presence of Si–H modes have also been identified in the silicon suboxide matrix prepared under certain conditions. CW and time-resolved photoluminescence (PL) measurements have exibited a broad band emission ranging from the UV to mid-infrared region with decay times ranging from picosecond to nanosecond time scales. CW PL reveals an emission at around 700 nm, which is well correlated with the oxide layer thickness. The spectroscopic ellipsometry provided us with the identification of the energy band structure through the presence of the critical point energy transitions. I–V characteristics of MOS devices are indicative of a resistive switching and electro-formation mechanism in the layers. The results have been anlyzed using advanced correlation techniques.

The manuscript provides a comprehensive summary of the development in this area in an attempt to evaluate the feasibility potential of silicon suboxide layer so grown for advanced applications ranging from anodes in lithium batteries to optical components[1]. Silicon suboxide (1 < x < 2) has been attracted for its abundance, low cost, easy synthesis, ideal gravimetric capacity and relatively small volume expansion to be used as anode[13].

## Experimental methods
### Growth conditions

A number of techniques have been used to prepare silicon suboxide layers, including thermal or electron beam evaporation, laser ablation and sputtering[14–17]. Among others, plasma enhanced chemical vapor deposition (PECVD) occupies a special place in preparing application quality SiOx layers[18]. Compared with these techniques, our method of suboxide preparation offers a quite suitable method for practical large area coverage and cost-effective production[12]. Moreover, it is compatible for integrated circuit fabrication processes since it is derived from the silicon wafer. Still our technique is far beyond the traditional methods for producing SiOx with good quality control in terms of the composition and the thickness of the layers.

Silicon suboxide layer was formed during the growth of $(NH_4)_2SiF_6$ crystals when Si wafer is exposed to the vapor of $HNO_3$:HF chemical mixture[12]. It is likely that this exposure induces oxidation on the wafer surface following the reaction $4Si + 6HF + 2HNO_3 \rightarrow (NH_4)_2SiF_6 + 3SiO_x$. But it is also required for the growth of $(NH_4)_2SiF_6$, hence the name oxide layer is due. Figure 1 shows the cross-sectional image of such a growth process and the energy dispersive X-ray or spectroscopy (EDX or EDS from SEM JEOL-JSM-6335S) measurements to determine the concentration distribution of elements. Ammonium silicon hexafluoride crystal layer is about 100 μm thick as a result of several hours of growth, the oxideoxide layer has almost fixed thickness of about sub-micron. Also shown is the distribution of elements throughout the cross section from the surface to the wafer. The green curve is typical density distribution of Si atoms, picking at the interface around the oxide layer. This peak is followed by a dip within the wafer and then reaches a constant value deep inside wafer, which could be due to silicon consumption during the $(NH_4)_2SiF_6$ formation. The insert indicates the fluorine concentration, which is relatively high in the $(NH_4)_2SiF_6$ layer as expected (yellow color). The purple and blue curves are Aluminum (Al) and Tantalum (Ta), respectively. These metals are due to surface metallization but their concentration is almost negligible. From the weighted ratio, the X was determined to be ranging from 1.2 up to 0.12 in silicon sub-oxide layer, SiOx. Also, inserted in Fig. 1, a high resolution transmission electron microscopy (TEM, JEM 4010) image of the interfacial layer between the Si and the dielectric layer. The image exhibits some porous structure with void like features and some pore like formations. The overall image represents a typical silicon surface structure having plenty of defects. This TEM image does not show any evidence for the presence of nanocrystalline formations.

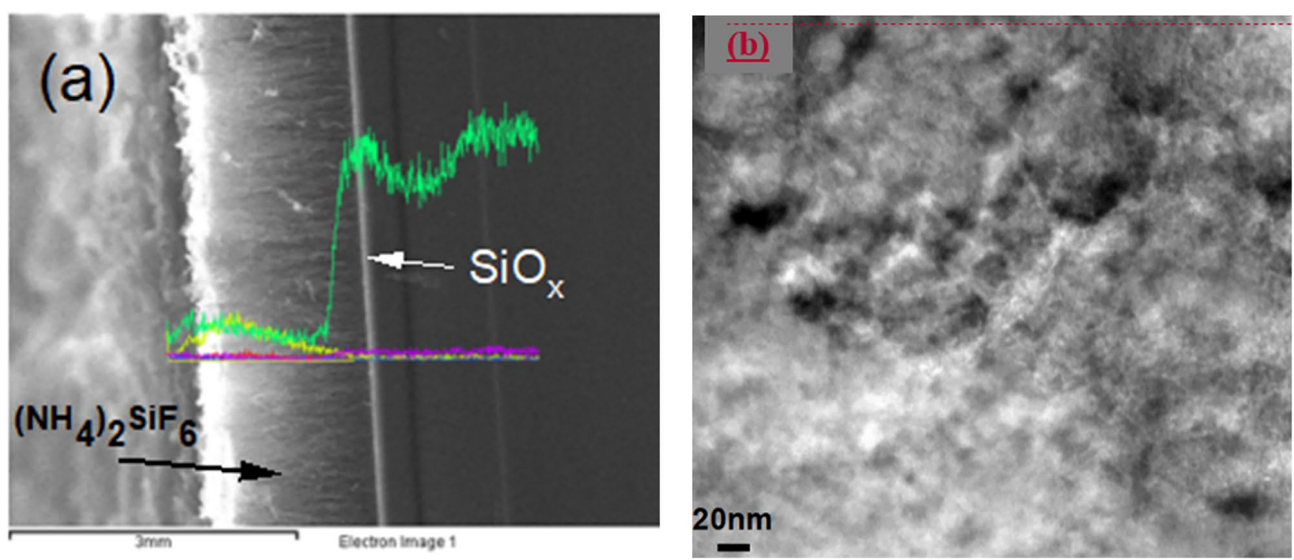

**Figure 1.** (**a**) Energy dispersive X-ray (EDX) spectrum taken at SEM from the cross-section of the sample indicating the distribution of elements. Colored line indicates Silicon (green line), Fluorine (yellow line), Aluminum (Purple), Nitrogen (Red) and Tantalum (blue). (**b**) Also, inserted a TEM image of the Oxide interface with 20 nm of magnification.

### Scanning probe analysis (atomic force microscopy)

Results of the Atomic Force Microscopy, AFM, measurements are shown in Fig. 2 indicating the surface topography of suboxide layer. The oxide layer with the scan area of $1 \times 1$ μm$^2$, prepared with 10% HF and 42% HNO$_3$ for 4 h indicate a porous-like surface structure. In Fig. 2b, the height histogram indicating a roughness of RMS = 33.9 Å and the bearing curve indicates a relatively smooth surface with occurrence of peaks and valleys.

The bearing curve gives the areal material ratio under the profile at any given depth as compared to a perfectly flat profile. The parameters marked on the curve indicate the core roughness $R_c$ (50 nm), reduced peak height $R_p$ (40 nm) and the reduced valley depth $R_v$ (40 nm), respectively.

The experimental methods also include Fourier transformed infrared spectroscopy (FTIR), Raman and photoluminescence using Argon ion laser for excitation, spectroscopic ellipsometry (SE) (Woolam) and current–voltage measurements (Keithley 4200 Semiconductor Parameter Analyzer) at room temperature.

### FTIR spectroscopy: Si–O–Si stretching modes

The localized vibrational modes related to Si–O–Si stretching frequencies are represented in Fig. 3. For oxides grown on plane wafer, the frequency of these modes appear at around 1100 cm$^{-1}$ as a doublet at 1081 cm$^{-1}$ and 1185 cm$^{-1}$ (solid red line). The spectrum is compared to that of DI water rinsed sample having the lower frequency mode shifted by 5 cm$^{-1}$ and the higher frequency is smoothed (solid black line, Fig. 3a).

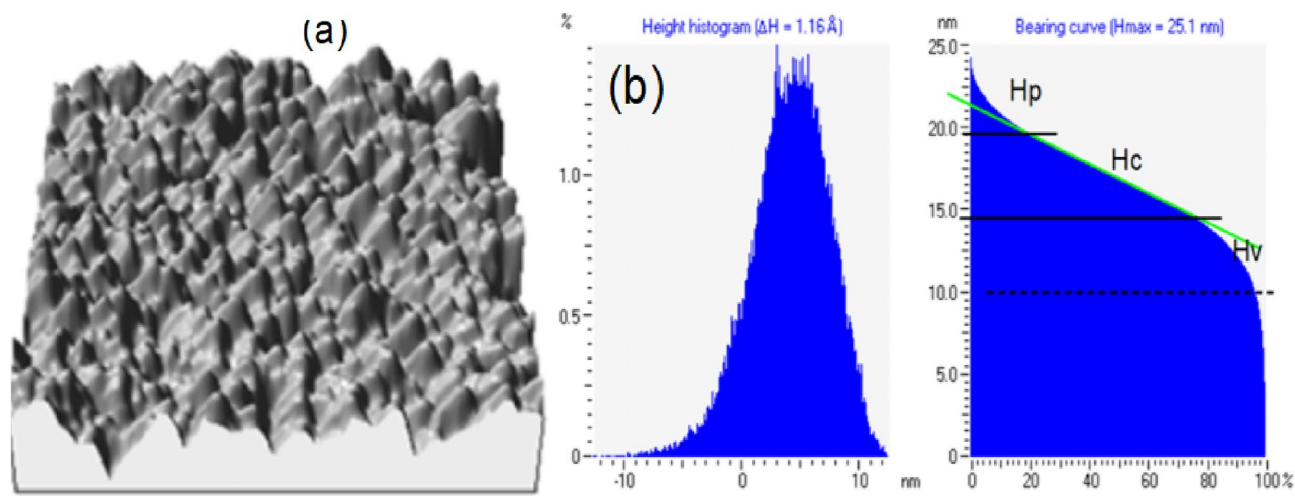

**Figure 2.** Scanning probe microscopy analysis. (**a**) 3D AFM topography image of the oxide layer with the scan size of $1 \times 1$ μm$^2$, prepared with 10% HF and 42% HNO$_3$ for 4 h. (**b**) The height histogram indicating roughness of RMS = 33.9 Å and the bearing curve. Where $R_c$, $R_p$ and $R_v$ are the core roughness, reduced peak height and reduced valley depth, respectively.

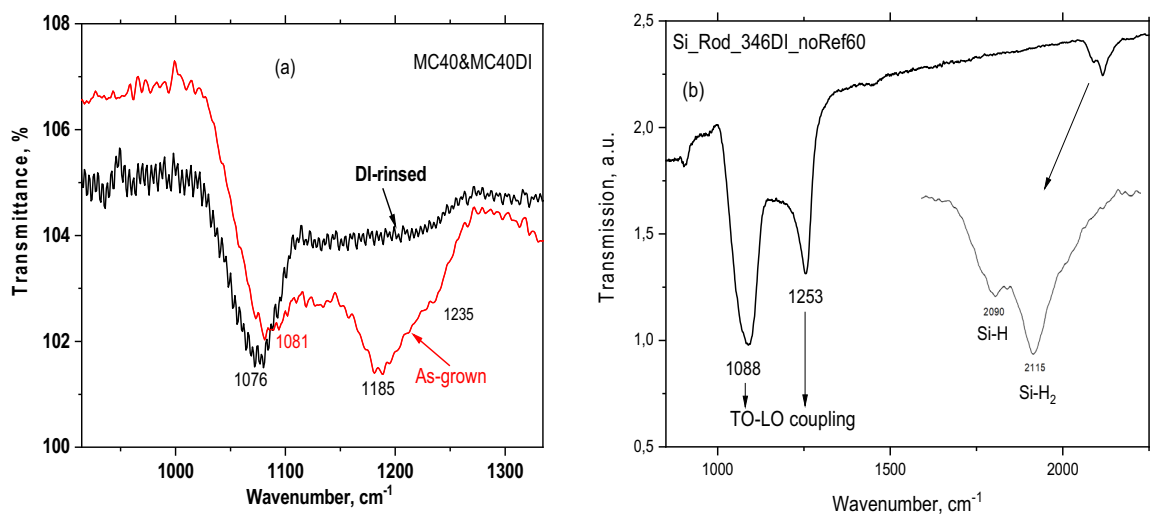

**Figure 3.** FTIR spectra indicating vibrational modes of Si–O–Si and Si–H bonds. (**a**) Si–O–Si stretching vibrations of an as grown oxide compared with the same sample after DI water rinsing. (**b**) Oxide vibrations measured on Si wafer consisting of Si rods of 400 nm in diameter. The insert indicates the presence of Si–H and Si–H$_2$ vibrational modes.

In Fig. 3b, we observe the splitting of Si–O–Si vibrational bands due to transverse-optic (TO) and longitudinal-optic (LO) modes coupling[19,20]. The forming rods on Si surface increases the surface area, which in turn enhances the splitting of LO-TO coupling modes. The vibration that we observe at 1076 cm$^{-1}$ belongs to optically active oxygen asymmetric stretch TO mode (in-phase motion of adjacent oxygen atoms). The coupling can be experimentally measured by in oblique incidence p-polarized transmission spectrum as peaks at around 1076–1253 cm$^{-1}$ on Si wafers[20]. The larger the coupling the higher the structural disorder in the oxide.

Another important feature of the spectrum is the incorporation of Hydrogen atoms into SiOx matrix as evidenced from the FTIR spectrum in Fig. 3b, at around 2000–2200 cm$^{-1}$. These vibrations are attributable to Si-H$_n$ (n = 1, 2, 3) modes[20]. Note that this incorporation occurs only under certain preparation conditions such as thicker oxide and when oxide is grown on silicon rods. This is to increase the strength of the absorption bands enabling us to resolve the hidden bands such as Si–H, otherwise not visible in thin oxide layers. These rods having much larger surface area are made of crystalline silicon but covered also by SiOx layer using the same method. For the same reason, TO–LO mode coupling becomes apparent. Tsu et al.[21] using local bonding environments of Si–O stretching groups modeled the stretching band as a function of alloy composition in silicon suboxides grown by PECVD. The expression $\nu = 965 + 50 \times$ cm$^{-1}$ is used to find x from experimentally determined position of the peak frequency. However, Zamchiy et al.[22] have found a significant overestimation of the x values obtained from Si–O–Si stretching mode position in the FTIR spectra of the films deposited by gas-jet electron beam plasma CVD. Instead they used the integrated absorption of the related band and related it to the atomic density N$_O$ of Oxygen atoms

$$N_o = A \cdot I = A \cdot \int \frac{\alpha(\omega)}{\omega} d\omega,$$

where A and I is the proportionality coefficient A (cm$^{-2}$) and the integral expression, respectively. N$_o$ is the atomic density of Oxygen N$_O$ (cm$^{-3}$). A widely used method to estimate x value in silicon suboxide films lies on the measurement of the wavenumber of the Si–O–Si stretching peak position that is in a linear relationship with x according to Ref.[21] $x = 0.020\nu_{st} - 19.3$. However, an overestimation of x values were obtained using this approximation. We find that the Raman stretching modes provides more reasonable values. Also, the band gap and thus the photoluminescence peak position has been correlated with the amount of the chemical composition x[23]. Based on this correlation, the observed photoluminescence peak at room temperature would correspond x values of about 1.3 to 1.5.

### Light emission

*Continuous wave (CW) photoluminescence (PL)*
CW PL measurements have been performed at room temperature and at low temperature 10 K using an Ar ion laser of 514 nm. The results are shown in Fig. 4 for temperature and surface condition dependent behavior. Using the equation of the Gauss model, photoluminescence emission curve can be fitted by the following equation, resulting in peaks at around 690 nm, 630 nm and 590 nm. Broad band peaked at around 608 nm can be deconvoluted into three emission bands peaked at 689 nm, 636 nm and 589 nm. A number of studies have carried out on Si nanocrystals and the origin of the emission peaks from about the Si band-gap (1.1 eV) to higher energies have been attributed to quantum confinement effects[18,24]. These PL emissions from the various size nanocrystals were named as S-band emission and our peaks at 1.8 eV (689 nm), 1.95 eV (636 nm) and 2.1 eV (589 nm) would correspond to nanocrystals with average diameter of 2.6 nm, 2.3 nm and 1.8 nm, respectively. However, we cannot rule out the effect of oxygen related defects for the origin of the light emission in the same region.

The shape and the location of the PL emission bands vary depending on the preparation conditions. For example, the band at 615 nm at 300 K in Fig. 4b shifts to higher energies, 2.1 eV at the low temperature, 10 K. Also, a broadening of 40% in low temperature PL was observed with respect to 300 K spectrum. This broadening can be attributed to the activation of additional nanocrystals as the temperature is lowered. Another interesting feature is the initial steps of the oxide formation as shown in Fig. 4c, where the PL peaks at 740 nm for 1 min exposure time and shifts to higher energy, that is 735 nm for 80 s of exposure. A second band appears also as a shoulder at 770 nm at short exposure time. The results indicate that the PL emission bands moves to higher energies with the processing time. Figure 4d summarizes the incidences of observed main PL peak positions for a number of samples.

*Time-resolved light emission properties*
The temporal and spectral dependence of the SiOx oxide layer are shown in Fig. 5a,b. The PL emission was obtained at 300 K between 470 and 600 nm using 400 nm (3.1 eV) excitation line of a Ti–Sapphire laser with a power levels up to 8 mW. The intensity distribution was deduced by steak camera image.

The temporal curves have been fit using a simulation program resulting in three components lifetimes as typical dynamical behavior as shown in Fig. 5a. The following expression was found to be well describing the luminescence decay time

$$I = \sum_i A_i \exp\left(\frac{t}{\tau_i}\right),$$

where i = 1, 2, 3. A$_i$ is a constant and $\tau$ is the lifetime of carriers.

The results of this analysis reveal three different lifetimes or decay components: 8.9 ps, 37.5 ps and 4500 ps, which can be attributed to ultrafast recombination of carriers at nanostructured Si, at silicon-oxide interface and oxide color centers, respectively. However, we don't exclude the attribution of these emissions originating from nanostructured silicon of different sizes leading to the quantum confinement of carriers. By analogy to previous studies, the short lived states related emission is due smaller size crystals and the long lived states could be attributed to larger crystals[25].

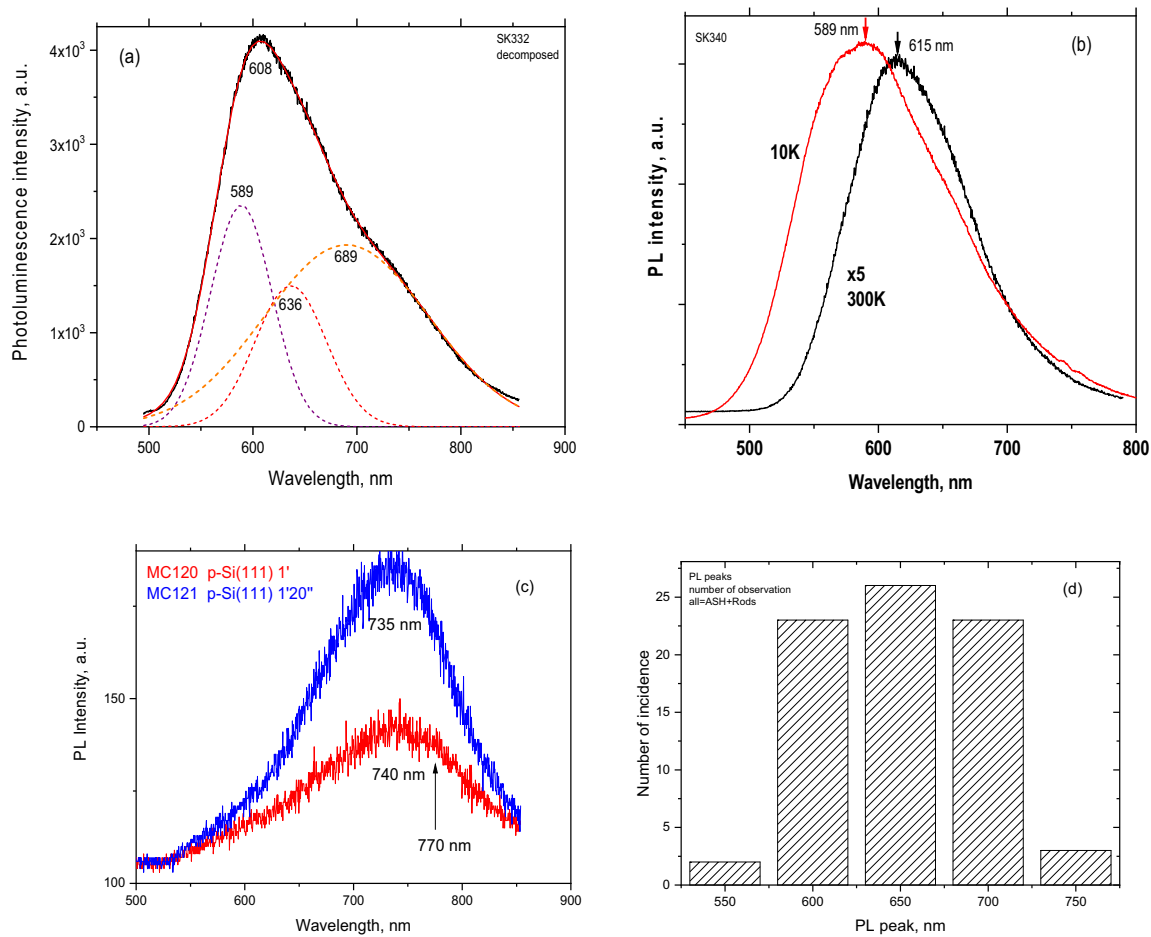

**Figure 4.** CW photoluminescence of SiOx oxide layer measured under different surface conditions. (**a**) The plane surface, (**b**) temperature effect, (**c**) initial oxide growth, (**d**) incidence of PL peaks on a number of SiOx oxide sample.

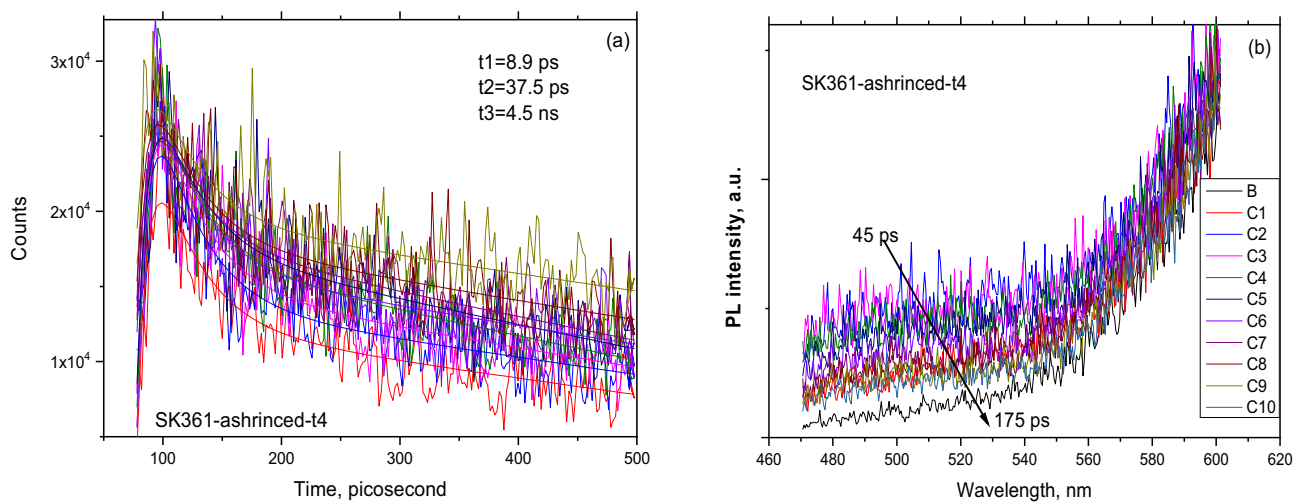

**Figure 5.** (**a**) Temporal and (**b**) spectral characteristics of the time-resolved photoluminescence measurements.

The time dependence of the spectral distribution of the emission is shown on Fig. 5b from 45 to 1975 ps time frame. The emission intensity decreases with time at shorter wavelengths, indicating the charge transfer to long lived states, that is red emission region.

*Spectroscopic ellipsometry (SE)*
The spectroscopic ellipsometry measurement parameters Ψ and Δ are shown in Fig. 6a and the calculated dielectric functions are plotted in Fig. 6b. Moreover, the second order derivative of spectroscopic ellipsometry measurementindicate interband critical electronic energy points, E1 (3.4 eV) E2 (4.3 eV). These critical energy points are typical of crystalline silicon, hence originating from the wafer[5]. The features at higher energies, that is 4.6 eV and 4.8 eV are rather attributable to some structural disorder and irregularities, if not quantum confinement effect at smaller size silicon nanostructures.

The SE of the initial stages of the SiOx formation samples exhibit sharp peaks at relatively lower energies, 1.35 eV and 1.65 eV, which might correspond to nanocrystal sizes of 5.4 nm and 2.5 nm, respectively, if we considered a quantum confinement effect[25].

Spectroscopic ellipsometry measures the amplitude ratio **ρ** of the reflection coefficient $r_p$ and $r_s$ (**p**-parallel and **s**-perpendicular polarizations to the plane of incidence, respectively). This can be expressed in terms of the amplitude ratio **tan ψ** and the phase angle **Δ**, which are the measured SE parameters as shown in Fig. 6a,b:

$$\rho = \frac{r_p}{rs} = (\tan\psi)e^{(i\Delta)},$$

where ρ is the ellipsometric ratio; Ψ and Δ denote the amplitude ratio and the phase difference, respectively; $r_p$ and $r_s$ are the Fresnel reflection coefficients in *p*- and *s*-polarizations. This dependence on the amplitude ratio and the phase difference between p- and s-polarizations indicates that the SE parameters are very sensitive to the surface roughness.

From the AFM, SEM and HRTEM, we observe that the SiOx layer has some surface roughness and void/porosity related formations. This type of intermixing of mediums can be best simulated using Bruggeman effective media approximation (EMA) to assign the optical constants of the layer[26,27]. The complex dielectric function

$$\varepsilon(\omega) = \varepsilon_1(\omega) + i\varepsilon_2(\omega),$$

can be derived by combining imaginary and reel parts using Fresnel equations, where the complex ε is expressed as follows

$$\varepsilon = \sin^2\varphi\left[1 + \tan^2\varphi\frac{(1-\rho)^2}{(1+\rho)^2}\right],$$

where **φ** is the incident angle of light. The real $\varepsilon_1(\omega)$ and the imaginary $\varepsilon_2(\omega)$ components of the dielectric functions so calculated are shown in Fig. 6. The spectra are dominated by a constant loss band up to 1000 nm and the negative **ε₁** values in the near UV–Vis. The dielectric contrast between the matrix and the nanostructures can be evaluated by using $\frac{\varepsilon_2}{\varepsilon_2-\varepsilon_1}$[28], indicating that it is the lowest at high energies and decreases down to band gap energies.

To evaluate the volume fractions of the constituents, one can use Bruggeman effective medium approximation (BEMA)[26,29], that is defined as

$$\sum_i f_i \frac{\varepsilon_i - \varepsilon_{eff}}{\varepsilon_i + 2\varepsilon_{eff}} = 0,$$

where $f, \varepsilon_i, \varepsilon_{eff}$ is the volume fraction and dielectric function of $i$th component and the effective dielectric function of the medium, respectively. Assuming a porous layer consisting of an amorphous Si and pores (voids) the above equation can be expressed as

$$f \frac{\varepsilon_v - \varepsilon_{eff}}{\varepsilon_v + 2\varepsilon_{eff}} + (1-f) \frac{\varepsilon_{Si} - \varepsilon_{eff}}{\varepsilon_{Si} + 2\varepsilon_{eff}} = 0,$$

where $\varepsilon_v$ and $\varepsilon_{Si}$ are the dielectric functions of the voids/pores and the dielectric constant of Si. The symbol $f$ represents the porosity of the layer. The dielectric constant for voids $\varepsilon_v$ can be assumed as equal to 1 and $\varepsilon_{Si}$ is taking from the literature as 12[30–32]. Assuming $\varepsilon_{eff} = 2.7$ from the data, the volume fraction of pores was estimated to be about 20%.

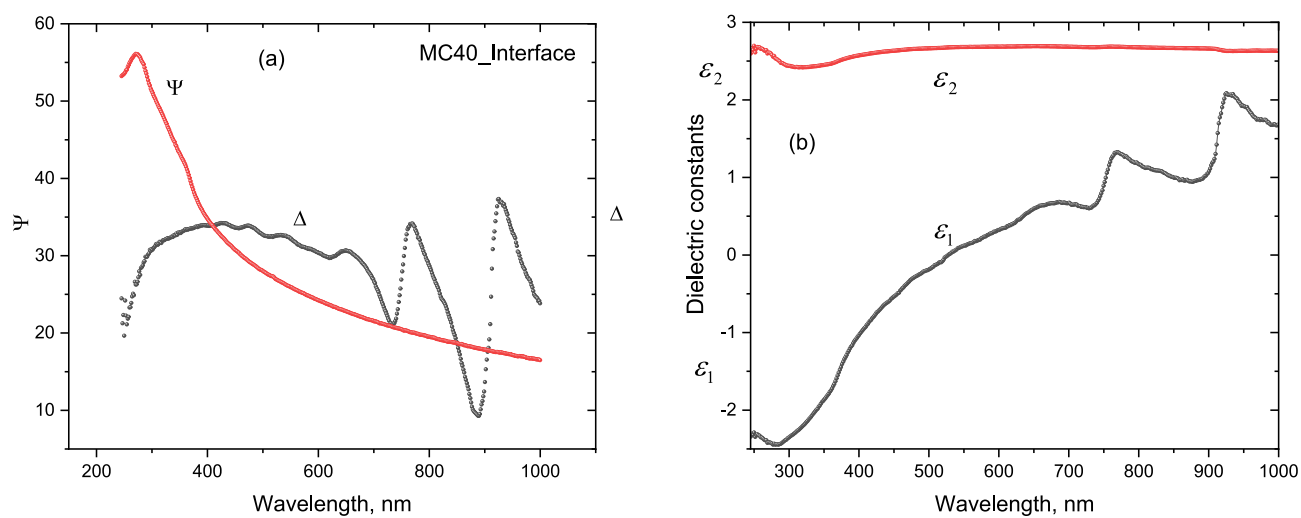

**Figure 6.** Spectroscopic ellipsometry at 300 K. (**a**) Ψ and Δ denote the amplitude ratio and the phase difference of measured SE parameters of the layer (**b**) Calculated real **ε**₁ and imaginary **ε**₂ dielectric functions of the suboxide layer.

*Resistive switching properties*
Resistive switching characteristics of the SiOx oxide layer are shown in Fig. 7 on consecutive voltage sweeps. I–V measurements were performed using Keithley 4200 Semiconductor Parameter Analyzer system and a probe station at room temperature. The measurements were done on devices with 50 nm Pt top electrode. Silicon wafer was used as bottom electrode.

As shown in Fig. 7a, the forming starts at around 4 V with a double forming steps for this particular sample. Upon consecutive sweeps, the same characteristic was reproduced. The saturation of the current at higher voltages is due to the compliance current set at 150 mA. Resistive switching or forming in silicon suboxide was found to be an intrinsic property of the suboxide. It has been proposed that the switching mechanism is driven by competing field-driven formation, and current-driven switch-off the conductive filament[30,33,34]. The switchable conductive pathway was demonstrated to be due to trap assisted tunneling through silicon nanocrystals embedded within the oxide layer.

Figure 7b is the current–voltage characteristics indicating the increase in conductance after successive SET process, indicating multi-step switching effect. This behavior can be attributed to the fact that the forming process is not completed at this voltage range explored[35]. The current conduction mechanism can be attributed to a thermally induced hopping. Similar behavior has been observed in a number of materials exhibiting resistive switching properties[36]. In amorphous silicon suboxides, it was reported that the presence of columnar microstructure is a key factor for the resistive random switching. The higher the RMS value of the surface roughness the higher the electroforming voltage. Thus the interface quality of the suboxide and the electrode plays a significant role in the switching properties[37]. We attribute relatively high threshold values to thick wafer effect. We conclude that much better switching performance could be observed when the suboxide oxide layer is grown on SOI wafers.

## Conclusion
A comprehensive description of the physical and electrical properties of the silicon suboxide oxide layer formed during the ammonium silicon hexafluoride crystals was provided in this work. The relatively rough surface structure as determined by scanning probe analysis provided 3D image of the nanostructured morphology with an RMS value of few nanometers. The presence of TO–LO coupling of asymmetric stretching modes of Si–O–Si bonding and the activation of this coupling was clearly demonstrated. The sample preparation conditions can be arranged to incorporate hydrogen atoms into the suboxide matrix as evidenced by localized Si–H vibrational modes.

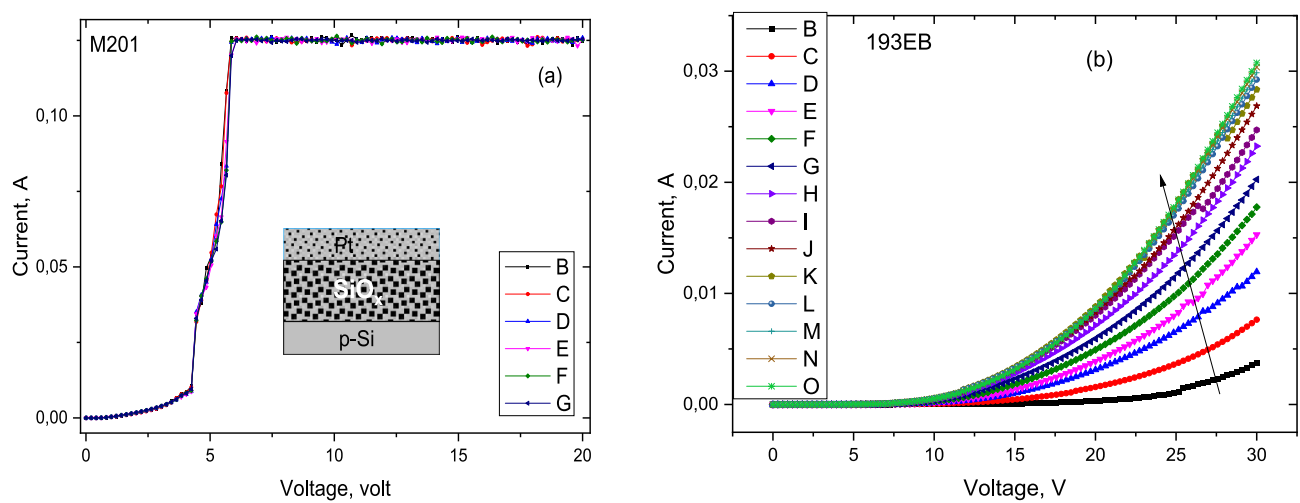

**Figure 7.** Resistive switching properties: analog switching characteristics on consecutive voltage sweeps. (**a**) The electroforming in a Pt/SiOx/p-Si device. (**b**) Under positive sweep indicating the increase of the conductance in SET process.

The suboxide layer exhibits an efficient photoluminescence emission in the region from blue to 900 nm with an average peak PL energy at around 650 nm but with a shift down to 500 nm and up to 800 nm. The low temperature PL indicates an enhancement in the emission intensity, which accompanied by a blue shift. The emission bands can be associated with the oxide related defects or quantum sized silicon crystals. The critical energy points as deduced from spectroscopic ellipsometry are compatible with the defect emission and quantum confinement effects.

The lifetime measurements are indicative of ultrafast components in the low range of picosecond regime, which could be associated with the radiative recombination in the smallest size nanostructures, in which electron and hole wave functions are most overlapped. Also, a charge transfer to long lived states in defects enhances red emission with a longer decay time.

Concerning resistive switching properties, the presence of forming is well observed, suggesting the possibility of application as resistive non-volatile memories, ReRAM. We think that the suboxide switching would perform best when the device is fabricated on a SOI wafer, in which the thinner silicon layer would favor lower threshold levels for switching. Using the method described in this paper, Si suboxide can be prepared on any micro and nanostructured silicon surface including nanowires, nanodots, rods, any artificial geometrical surface and plane surface depending on application requirements. It is also suitable for industrial production on large silicon wafer area selectively with a large quantity.

## Data availability
The datasets generated during and/or analyzed during the current study are available from the corresponding author on a reasonable request.

## References
1. Fricke-Begemann, T., Meinertz, J., Weichenhain-Schriever, R. & Ihlemann, J. Silicon suboxide (SiOx): Laser processing and applications. *Appl. Phys. A* **117**, 13 (2014).
2. Kalem, S. 1320 nm light source from deuterium treated silicon. *IEEE Open J. Nanotechnol.* **1**, 88 (2020).
3. Kissinger, G. *et al.* Precipitation of suboxides in silicon, their role in gettering of copper impurities and carrier recombination. *ECS J. Solid State Sci. Technol.* **9**, 064002 (2020).
4. Ugwumadu, C. *et al.* Structure, vibrations and electronic transport in silicon suboxides: Application to physical unclonable functions. *J. Non-Cryst. Solids X* **18**, 100179 (2023).
5. Kong, X. *et al.* Recent progress in silicon-based materials for performance-enhanced lithium-ion batteries. *Molecules* **28**, 2079 (2023).
6. Zhang, M. *et al.* Recent advances of SiOx-based anodes for sustainable lithium-ion batteries. *Nano Res. Energy* **2**, e9120077 (2023).
7. Editorial. Reading through breakthroughs. *Nat. Energy* **2**, 17126 (2017).
8. Frith, J. T., Lacey, M. J. & Ulissi, U. A non-academic perspective on the future of lithium-based batteries. *Nat. Commun.* **14**, 420 (2023).
9. Transparency Market Research. *Silicon Based Anode Market, Market Forecast Value in 2031 is 3.4 Bn USD*. https://www.transparencymarketresearch.com/silicon-based-anode-market.html.
10. BIS-Research. *Global Next Generation Anode Materials Market*. https://bisresearch.com/industry-report/next-generation-anode-materials-market.html (2022).
11. Liu, H. *et al.* The application road of silicon-based anode in lithium-ion batteries: From liquid electrolyte to solid-state electrolyte. *Energy Storage Mater.* **55**, 244 (2023).
12. Kalem, S. Synthesis of ammonium silicon fluoride cryptocrystals on Silicon. *Appl. Surf. Sci.* **236**, 336 (2004).
13. Zhou, X. *et al.* Research progress of silicon suboxide based anodes for lithium ion batteries. *Front. Mater.* **7**, 628233 (2021).
14. Hong, S. H. *et al.* Optical characterization of Si nanocrystals in Si-rich SiOx and SiOx/SiO2 multilayers grown by ion beam sputtering. *J. Korean Phys. Soc.* **45**, 116 (2004).
15. Kahler, U. & Hofmeister, H. Visible light emission from Si nanocrystalline composites via reactive evaporation of SiO. *Opt. Mater.* **17**, 83 (2001).
16. Iacona, F., Franzo, G. & Spinella, C. Orrelation between luminescence and structural properties of Si nanocrystals. *J. Appl. Phys.* **87**(3), 1295–1303 (2000).
17. Starinskiy, S. V., Rodionov, A. A., Shukhov, Y. G. & Bulgakov, A. V. Oxidation of ablated silicon during pulsed laser deposition in a background gas with different oxygen partial pressures. *EPJ Web Sci.* **196**, 00008 (2019).
18. Tsu, D. V., Kim, S. S., Theil, J. A., Wang, C. & Lucovsky, G. Formation of multilayer SiO2-SiOx heterostructures by control of reaction pathways in remote PECVD. *MRS Proc.* **165**, 1 (1989).
19. Kirk, C. T. Quantitative analysis of the effect of disorder-induced mode coupling on infrared absorption in silica. *Phys. Rev. B* **38**, 1255 (1988).
20. Kalem, S. Infrared spectroscopy of hydrogenated and chlorinated amorphous silicon. *Philos. Mag.* **53**, 509 (1986).
21. Tsu, D. V., Lucovsky, G. & Davidson, B. N. Effects of the nearest neighbors and the alloy matrix on SiH stretching vibrations in the amorphous SiOx alloy system. *Phys. Rev. B* **40**, 1795 (1989).
22. Zamchiy, A. O. *et al.* Determination of the oxygen content in amorphous SiOx thin films. *J. Non-Cryst. Solids* **518**, 43 (2019).
23. Ming, T. *et al.* Suboxides as driving force for efficient light. *Small* **17**, 2007650 (2021).
24. Canham, L. Origins and applications of efficient visible photoluminescence from silicon based nanostructures. *R. Soc. Chem. Faraday Discuss.* **222**, 10 (2020).
25. Petrik, P. *et al.* Nanocrystal characterization by ellipsometry in porous silicon using model dielectric function. *J. Appl. Phys.* **105**, 2 (2009).
26. Aspnes, D. E. In *Handbook of Optical Constants of Solids* (ed. Palik, E. D.) 104 (Academic, 1985).
27. Kanneboina, V. & Agarwal, P. Spectroscopic ellipsometry investigation to study microstructure evolution in B-doped a-Si. *SN Appl. Sci.* **3**, 500. https://doi.org/10.1007/s42452-021-04495-7 (2021).
28. Keita, A. S. *et al.* Determination of optical properties and size dispersion of Si nanoparticles within a dielectric matrix by SE. *J. Appl. Phys.* **116**, 103520 (2014).
29. Sachse, R. *et al.* Multilevel effective material approximation for modelling ellipsometric measurements on complex porous thin films. *Adv. Opt. Technol.* **11**, 137 (2022).
30. Fodor, B. *et al.* Spectroscopic ellipsometry of columnar porous Si thin films and nanowires. *Appl. Surf. Sci.* **421**, 397 (2017).

### Acknowledgements
This work did not receive any specific grant from funding agencies. It was supported by European Commission LASERLAB-EUROPE program (LLC001765) and MC2ACCESS for facility use. The authors thank Prof. W. Sunström for the access to Lund Laser Center and Dr.A.E. Hannas for the assistance in femtosecond laser measurements and Ö. Arthursson for AFM probing of surfaces. BMBF and TUBITAK are acknowledged for the bilateral program and Dr. P. Werner and Dr. V. Talalaev for TEM and low temperature photoluminescence measurements.

### Author contributions
Seref Kalem designed the experiments, has grown suboxide, analyzed the results and wrote the manuscript.

### Competing interests
The authors declare no competing interests.

### Additional information
**Correspondence** and requests for materials should be addressed to S.K.

**Reprints and permissions information** is available at www.nature.com/reprints.

**Publisher's note**  Springer Nature remains neutral with regard to jurisdictional claims in published maps and institutional affiliations.

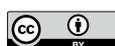